\documentclass[aps,pra]{revtex4}
\usepackage{amssymb}
\usepackage{graphicx}
\usepackage{amsmath}
\usepackage[sort&compress]{natbib}
\usepackage{color}
\usepackage{bm}
\usepackage{epsfig}
\usepackage{subfigure}

\begin{document}

\date{\today}
\title{Creating solitons by means of spin-orbit coupling}
\author{Boris A. Malomed}
\affiliation{Department of Physical Electronics, School of Electrical Engineering,
Faculty of Engineering, and Center for Light-Matter Interaction, Tel Aviv
University, Tel Aviv 69978, Israel}

\begin{abstract}
This mini-review collects theoretical results predicting the creation of
matter-wave solitons by the pseudo-spinor system of Gross-Pitaevskii
equations (GPEs) with the self-attractive cubic nonlinearity and linear
first-order-derivative terms accounting for the spin-orbit coupling (SOC).
In one dimension (1D), the so predicted bright solitons are similar to their
well-known counterparts supported by the GPE in the absence of SOC.
Completely novel results were recently obtained for 2D and 3D systems: SOC
suppresses the collapse instability of the multidimensional GPE, creating
fully stable 2D ground-state solitons and metastable 3D ones of two types:
\textit{semi-vortices} (SVs), with vorticities $m=1$ in one spin component
and $m=0$ in the other, and \textit{mixed modes} (MMs), with $m=0$ and $%
m=\pm 1$ present in both components. With the Galilean invariance broken by
SOC, moving solitons exist up to a certain critical velocity, suffering
delocalization above it. The newest result predicts stable 2D
\textquotedblleft quantum droplets" of the MM type in the presence of the
Lee-Huang-Yang corrections to the GPE system, induced by quantum
fluctuations around the mean-field states, in the case when the
inter-component attraction dominates over the self-repulsion in each
component
\end{abstract}

\maketitle

\section{Introduction}

Atomic gases, cooled to temperatures below the threshold for the transition
into a quantum degenerate state (in particular, Bose-Einstein condensates
(BEC) in bosonic gases \cite{BEC-book1,Bagnato}), find an important
application as a testbed allowing simulation of various effects, which were
originally known in complex forms in condensed-matter physics, and may be
realized (or emulated) in a simple and clean form in ultracold quantum gases
\cite{emulator}. BEC also offers a way to reproduce diverse phenomena which
were previously discovered in optics \cite{Malomed}. The latter possibility
is essentially based on the similarity of the nonlinear Schr\"{o}dinger
equation (NLSE), which is the basic propagation equation in optics \cite{KA}%
, and the Gross-Pitaevskii equation (GPE), which is a universal model for
ultracold bosonic gases \cite{BEC-book1}. In the framework of the
similarity, the same cubic nonlinearity, which represents the Kerr term in
optics, represents collisional effects in atomic BEC.

In particular, a binary ultracold \emph{bosonic} gas, whose two-component
mean-field wave function is considered as a pseudo-spinor, may emulate
spinor effects in the \emph{fermionic} gas of electrons with spin $1/2$ ,
even if the true bosonic spin is zero. In this vein, great attention has
been drawn to the experimentally demonstrated \cite{Nature} BEC-based\
emulation of the spin-orbit coupling (SOC) in semiconductors, i.e., the
linear interaction between the electron's momentum and its spin. In terms of
the atomic gas, SOC is mapped into a linear interaction of the atoms'
momentum and the pseudospin \cite{Nature}-\cite{Zhai}. The mapping makes it
also possible to include the Zeeman splitting (ZS)\ between up- and
down-states of the electron, which is an important aspect of the solid-state
SOC, represented by an energy difference between the two components of the
BEC binary wave function \cite{Campbell}. In semiconductors, two fundamental
types of SOC\ are represented by the Dresselhaus \cite{Dresselhaus} and
Rashba \cite{Rashba} Hamiltonians, each one admitting the emulation in
ultracold gases.

While majority of experimental works on the emulation of SOC were carried
out in effectively one-dimensional (1D) settings, the realization of the SOC
in the quasi-2D BEC was reported too \cite{2D-experiment}, encouraging the
consideration of SOC-supported modes in multidimensional geometries, such as
2D {\cite{Sakaguchi14}-\cite{NJP} and 3D \cite{HP} solitons, }vortices \cite%
{vortex1}-\cite{Fetter}, skyrmions \cite{skyrmions2}{, etc. A crucial role
in the creation of these nonlinear modes belongs to the sign of the
nonlinearity. Thus far, SOC was realized in the }$^{87}$Rb gas with
repulsive interactions. Nevertheless, in other bosonic species which admit
the transition to the BEC, the sign can be switched to attraction, by means
of the Feshbach resonance \cite{Cornish,Feshbach}. In particular, the
application of this method to the condensate of $^{7}$Li \cite{Randy,Salomon}
and $^{85}$Rb \cite{Weiman} atoms has made it possible the creation of
effectively 1D solitons.

This mini-review is focused on recently studied schemes allowing the
creation of solitons in binary BEC\ systems featuring SOC between the
components. Such experimental results were not yet reported, but a number of
theoretical predictions provide an incentive for the development of the
experiment in this direction. While in the 1D setting the addition of SOC
does not lead to qualitative changes of the previously studied soliton
phenomenology, the predictions for the 2D and 3D geometries suggest
possibilities of the realization of essentially novel effects. Arguably, the
most noteworthy ones are unique possibilities to create absolutely stable 2D
\cite{Sakaguchi14} and metastable 3D \cite{HP} solitons in the
self-attractive BEC, in spite of the presence of the critical and
supercritical collapse, driven in the 2D and 3D space, respectively, by the
cubic self-attraction \cite{Berge,Fibich} (peculiarities of the collapse in
SOC systems were considered in \cite{Mardonov15}). For this reason, the
creation of stable multidimensional optical and matter-wave solitons is a
challenging problem, which has drawn a great deal of interest, see reviews
\cite{review,Malomed} and \cite{me,Mihalache}. The above-mentioned recent
results make an essential contribution towards the solution of this problem.
In particular, the mini-review includes the newest addition to the topic
\cite{Raymond}, namely the prediction of stable SOC solitons in the
effectively 2D setting stabilized, at arbitrarily large values of their
norm, by beyond-the-mean-field corrections, \textit{viz}., the
Lee-Huang-Yang (LHY) terms \cite{LHY}, which are generated by quantum
fluctuations around the mean-field states. Recently, it had been predicted
theoretically \cite{Petrov,Grisha} and demonstrated experimentally \cite%
{Let1,Let2,Ing} that the LHY effect readily stabilizes two-component 3D
solitons, in the form of \textquotedblleft quantum droplets" (QDs), against
the collapse (in the system which does not include SOC effects).

The rest of the mini-review is structures as follows: basic models of SOC
systems are formulated in Section II, main results for 2D and 3D solitons
are summarized in Section III, and the article is completed by Section IV.

\section{The basic models: systems of coupled GPEs}

The system of GPEs in the 2D space $\left( x,y\right) $ for the
pseudo-spinor (two-component) wave function $\left( \phi _{+},\phi
_{-}\right) $, which includes the self-attraction with the coefficient
scaled to be $1$, cross-attraction with relative strength $\gamma >0$,
linear SOC\ terms of the Rashba and Dresselhaus types with respective
coefficients $\lambda $ and $\lambda _{D}$, and ZS with strength $\Omega >0$%
, is written as \cite{Spielman}-\cite{Zhai}, \cite{we2}
\begin{eqnarray}
&&\hspace{-11mm}i\frac{\partial \phi _{+}}{\partial t}=-\frac{1}{2}\nabla
^{2}\phi _{+}-(|\phi _{+}|^{2}+\gamma |\phi _{-}|^{2})\phi _{+}+\left(
\lambda D^{[-]}\phi _{-}-i\lambda _{D}D^{[+]}\phi _{-}\right) -\Omega \phi
_{+},  \label{GPERD1} \\
&&\hspace{-11mm}i\frac{\partial \phi _{-}}{\partial t}=-\frac{1}{2}\nabla
^{2}\phi _{-}-(|\phi _{-}|^{2}+\gamma |\phi _{+}|^{2})\phi _{-}-\left(
\lambda D^{[+]}\phi _{+}+i\lambda _{D}D^{[-]}\phi _{+}\right) +\Omega \phi
_{-},  \label{GPERD2}
\end{eqnarray}%
with $D^{\left[ \pm \right] }\equiv \partial /\partial x\pm i\partial
/\partial y$. The GPE system conserves the Hamiltonian, momentum, and the
total norm, which is proportional to the number of atoms in the condensate:%
\begin{equation}
N=\iint (|\phi _{+}|^{2}+|\phi _{-}|^{2})dxdy\equiv N_{+}+N_{-}.  \label{N}
\end{equation}%
In scaled equations (\ref{GPERD1}) and (\ref{GPERD2}) the unit length
corresponds to distance $\sim 1$ $\mathrm{\mu }$m and $N=1$ is tantamount to
$\simeq 3\times 10^{3}$ atoms \cite{we2}. The spectrum of excitations
generated by the linearized version of Eqs. (\ref{GPERD1}) and (\ref{GPERD2}%
), for $\phi _{\pm }\sim \exp \left( i\mathbf{k}\cdot \mathbf{r}-i\mu _{\pm
}t\right) $, where $\mathbf{k}$ is the wave vector, contains two branches
\cite{NJP}:
\begin{equation}
\mu _{\pm }=\frac{k^{2}}{2}\pm \sqrt{(\lambda ^{2}+\lambda
_{D}^{2})k^{2}+4\lambda \lambda _{D}k_{x}k_{y}+\Omega ^{2}}.  \label{eps}
\end{equation}%
Solitons may exist at values of $\mu $ which are not covered by Eq. (\ref%
{eps}) with real wavenumbers, i.e., at%
\begin{gather}
\mu <-\frac{1}{2}\left[ \left( |\lambda |+\left\vert \lambda _{D}\right\vert
\right) ^{2}+\frac{\Omega ^{2}}{\left( |\lambda |+\left\vert \lambda
_{D}\right\vert \right) ^{2}}\right] ,~\mathrm{if}~~\left( |\lambda
|+\left\vert \lambda _{D}\right\vert \right) ^{2}>|\Omega |,  \notag \\
\mu <-|\Omega |,~\mathrm{if}~~\left( |\lambda |+\left\vert \lambda
_{D}\right\vert \right) ^{2}<|\Omega |~.  \label{mu}
\end{gather}

The 3D model was addressed in \cite{HP}, with SOC of the \textit{Weyl type}
\cite{Anderson}:
\begin{gather}
\left[ i\frac{\partial }{\partial t}+\frac{1}{2}\nabla ^{2}+i\lambda \nabla
\cdot \mathbf{\sigma }\right.   \notag \\
\left. +\left(
\begin{array}{cc}
|\phi _{+}|^{2}+\gamma |\phi _{-}|^{2} & 0 \\
0 & |\phi _{-}|^{2}+\gamma |\phi _{+}|^{2}%
\end{array}%
\right) \right] \left(
\begin{array}{c}
\phi _{+} \\
\phi _{-}%
\end{array}%
\right) =0,  \label{GPE}
\end{gather}%
where $\nabla $ is the 3D gradient, $\lambda $ is the SOC coefficient, and $%
\mathbf{\sigma }=\left( \sigma _{x},\sigma _{y},\sigma _{z}\right) $ is the
set of three Pauli matrices. Actually, the existence of metastable 3D
solitons \cite{HP} is a generic fact, which is also valid for other
particular forms of SOC.

LHY corrections to the mean-field theory are relevant in the case when the
self-interaction of each component is repulsive, while the self-trapping of
solitons is provided by attraction between the components \cite{Petrov}. The
2D reduction of the original 3D GPEs with LHY terms replaces Eqs. (\ref%
{GPERD1}) and (\ref{GPERD2})\ by the system which contains nonlinear terms
with a logarithmic factor \cite{Grisha,Raymond}:%
\begin{eqnarray}
i\frac{\partial \phi _{+}}{\partial t} &=&-\frac{1}{2}\nabla ^{2}\phi
_{+}+\lambda D^{[-]}\phi _{-}+g(|\phi _{+}|^{2}-|\phi _{-}|^{2})\phi _{+}+%
\frac{g^{2}}{4\pi }\left( |\phi _{+}|^{2}+|\phi _{-}|^{2}\right) \ln \left(
|\phi _{+}|^{2}+|\phi _{-}|^{2}\right) \phi _{+},  \notag \\
i\frac{\partial \phi _{-}}{\partial t} &=&-\frac{1}{2}\nabla ^{2}\phi
_{-}-\lambda D^{[+]}\phi _{+}+g(|\phi _{-}|^{2}-|\phi _{+}|^{2})\phi _{-}+%
\frac{g^{2}}{4\pi }\left( |\phi _{+}|^{2}+|\phi _{-}|^{2}\right) \ln \left(
|\phi _{+}|^{2}+|\phi _{-}|^{2}\right) \phi _{-},  \label{GPE2}
\end{eqnarray}%
where $g>0$ is nonlinearity strength, and only SOC of the Rashba type, with
coefficient $\lambda $, is included.

\section{Stable 2D and 3D solitons in the SOC system: semi-vortices (SVs)
and mixed modes (MMs)}

\textit{Semi-vortices}. Basic results for 2D solitons stabilized by SOC are
presented here, following works \cite{Sakaguchi14,we2} and \cite{Romanian},
where technical details can be found. First, Eqs. (\ref{GPERD1}) and (\ref%
{GPERD2}) admit stationary solutions of the SV type (also called
half-vortices \cite{rf:2}). In the absence of the Dresselhaus and ZS terms, $%
\lambda _{D}=\Omega =0$, SVs are built as bound states of zero-vorticity ($%
m_{+}=0$) and vortical ($m_{-}=1$) self-trapped components, as per the
following exact ansatz, written in terms of polar coordinates $\left(
r,\theta \right) $, with chemical potential $\mu $:
\begin{equation}
\phi _{+}\left( x,y,t\right) =e^{-i\mu t}f_{1}(r^{2}),~\phi _{-}\left(
x,y,t\right) =e^{-i\mu t+i\theta }rf_{2}(r^{2}),  \label{frf}
\end{equation}%
with functions $f_{1,2}\left( r^{2}\right) $ obeying equations
\begin{eqnarray}
&&\mu f_{1}+2\left[ r^{2}\frac{d^{2}f_{1}}{d\left( r^{2}\right) ^{2}}+\frac{%
df_{1}}{d\left( r^{2}\right) }\right] +\left( f_{1}^{~2}+\gamma
r^{2}f_{2}^{~2}\right) f_{1}-2\lambda \left[ r^{2}\frac{df_{2}}{d\left(
r^{2}\right) }+f_{2}^{~}\right] =0,  \notag \\
&&\mu f_{2}+2\left[ r^{2}\frac{d^{2}f_{2}}{d\left( r^{2}\right) ^{2}}+2\frac{%
df_{2}}{d\left( r^{2}\right) }\right] +\left( r^{2}f_{2}^{~2}+\gamma
f_{1}^{~2}\right) f_{2}+2\lambda \frac{df_{1}}{d\left( r^{2}\right) }=0.
\label{ff}
\end{eqnarray}%
Due to their symmetry, Eqs. (\ref{GPERD1}) and (\ref{GPERD2}) also give rise
to a mode which is a counterpart of SV\ (\ref{frf}) with vorticities $\left(
m_{+},m_{-}\right) =$ $\left( 0,1\right) $ replaced by $\left(
m_{+},m_{-}\right) =\left( -1,0\right) $. At $\gamma \leq 1$, SVs represent
the ground state (GS) of the system, i.e., an absolute minimum of the energy
possible for a given norm (see details below). The coexistence of SV (\ref%
{frf}) and its counterpart implies \emph{degeneracy of the GS}, which is
possible in nonlinear systems, unlike linear ones. Solitons are solutions to
Eq. (\ref{ff}) localized at $r\rightarrow \infty $ as $\exp \left( \sqrt{%
-(2\mu +\lambda ^{2})}r\right) $, which exist at $\mu <-\lambda ^{2}/2$. In
the general case, with $\lambda _{D},\Omega \neq 0$, solitons exist in the
range of $\mu $ given by Eq. (\ref{mu}).

\begin{figure}[b]
\begin{center}
\includegraphics[width=0.7\textwidth]{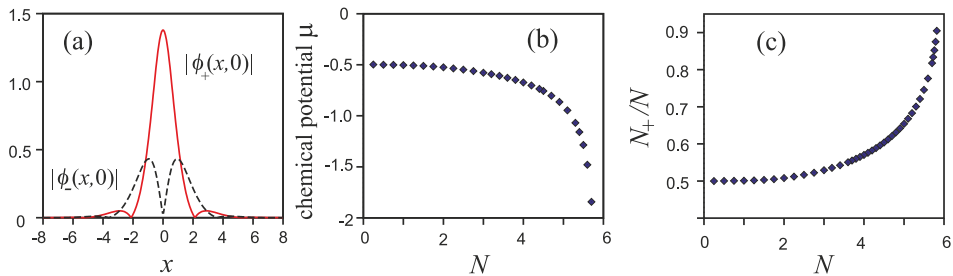}
\end{center}
\caption{(Color online) (a) Cross-sections of two components of the SV
(semi-vortex), $\left\vert \protect\phi _{+}(x,0)\right\vert $ and $%
\left\vert \protect\phi _{-}(x,0)\right\vert $, with norm $N=5$ and
parameters $\protect\lambda =1$, $\protect\gamma =\protect\lambda %
_{D}=\Omega =0$ in Eqs. (\protect\ref{GPERD1}) and (\protect\ref{GPERD2}).
(b) and (c) Chemical potential $\protect\mu $ and the relative share of the
norm in the zero-vorticity component, $N_{+}/N$ (see Eq. (\protect\ref{N})),
vs. $N$ for the SV family. The plots are borrowed from \protect\cite%
{Sakaguchi14} and \protect\cite{Romanian}.}
\label{fig1}
\end{figure}

A typical example of stable SVs, found as a numerical solution of Eq. (\ref%
{ff}), is displayed in Fig. \ref{fig1}. Results for the SV family are
summarized in Figs. \ref{fig1}(b,c), which display the chemical potential as
a function of the norm. It is seen that the zero-vorticity component always
carries a larger share of the total norm. The negative local slope of the $%
\mu (N)$ curve in Fig. \ref{fig1}(b) implies that it satisfies the
Vakhitov-Kolokolov (VK) criterion \cite{VaKo,Berge,Fibich}, which is a
necessary, but not sufficient, condition for the stability of solitons
supported by self-attractive nonlinearities. The SV family exists at $%
N<N_{\max }\equiv N_{\mathrm{T}}\approx 5.85$, the latter value being the
collapse threshold, i.e., the norm of the 2D \textit{Townes solitons} (TSs)
\cite{Townes} generated by the single GPE in the absence of SOC. Indeed,
Fig. \ref{fig1}(c) shows that the vortex component $\phi _{-}$ vanishes
in the limit of $N\rightarrow N_{\mathrm{T}}$ ($\mu \rightarrow -\infty $),
hence in this limit SV degenerates into the unstable single-component TS. At
$N>N_{\mathrm{T}}$, solitons do not exist, as the norm exceeding the
threshold value gives rise to the collapse. On the other hand, there is no
minimum value of $N$ necessary for the existence of stable SVs.

It is relevant to mention that the fundamental reason for the instability of
TSs in the usual 2D NLSE is the invariance of this equation with respect to
a scaling transformation, due to which all TSs have a single value of the
norm, $N_{\mathrm{T}}$. The presence of the SOC terms in Eqs. (\ref{GPERD1})
and (\ref{GPERD2}) breaks the scaling invariance and lifts the norm
degeneracy, pushing the soliton's norm to $N<N_{\mathrm{T}}$. This makes the
destabilization of the TSs by the collapse impossible, as it may only be
initiated by $N\geq N_{\mathrm{T}}$.

\begin{figure}[tbh]
\begin{center}
\includegraphics[width=0.7\textwidth]{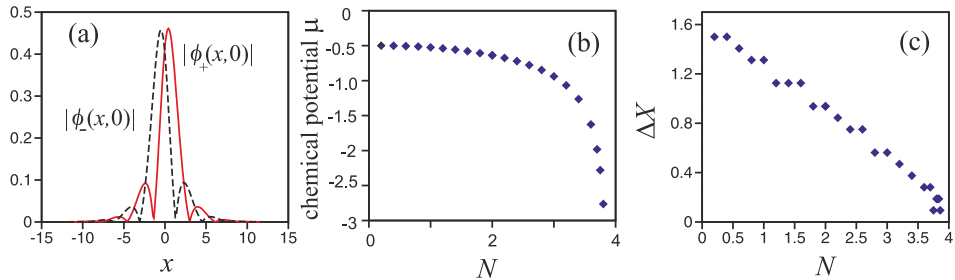}
\end{center}
\caption{(Color online) (a,b) The same as in Fig. \protect\ref{fig1} (a,b),
but for stable MMs (mixed modes) at $\protect\gamma =2$. (c) Separation $%
\Delta \,X$ between peak positions of $|\protect\phi _{+}(x,0)|^{2}$ and $|%
\protect\phi _{-}(x,0)|^{2}$ vs. the total norm. The plots are borrowed from
\protect\cite{Sakaguchi14} and \protect\cite{Romanian}.}
\label{fig2}
\end{figure}

\textit{Mixed modes}. Another type of 2D self-trapped vortical states
supported by Eqs. (\ref{GPERD1}) and (\ref{GPERD2}) can be produced by input
\begin{equation}
\left( \phi _{\pm }^{0}\right) _{\mathrm{MM}}=B_{1}e^{-\beta _{1}r^{2}}\mp
B_{2}r\,e^{\mp i\theta -\beta _{2}r^{2}},  \label{MM}
\end{equation}%
with real constants $B_{1,2}$ and $\beta _{1,2}>0$ (unlike the SV\ ansatz (%
\ref{frf}), this one is not compatible with the equations, being used only
an initial guess, or a basis for the variational approximation \cite%
{Sakaguchi14}). Modes generated by this input are called MMs as they mix
vorticities $\left( 0,-1\right) $ and $\left( 0,+1\right) $ in the two
components, which have equal norms (unlike SV). A typical example of the
stable MM and $\mu (N)$ dependence for the MM family, which again satisfies
the VK criterion, are displayed in Figs. \ref{fig2}(a) and (b), respectively
(for the system without the Dresselhaus and ZS terms, i.e., $\lambda
_{D}=\Omega =0$). The norms of the two MM's components are always equal,
while their maxima are separated by distance $\Delta X$, see Fig. \ref{fig2}%
(c). MMs exist at
\begin{equation}
N<N_{\max }\equiv 2N_{\mathrm{T}}/(1+\gamma ),  \label{N<N}
\end{equation}%
where $N_{\mathrm{T}}$ is the above-mentioned TS\ norm. In the limit of $\mu
\rightarrow -\infty $ ($N\rightarrow N_{\max }$),\ the vortex terms vanish
in the MM, and it degenerates into an unstable two-component TS, similar to
the above-mentioned degeneration of the SV. Other similarities are that $%
N>N_{\max }$ leads to the collapse, and, on the other hand, there is no
minimum norm necessary for the existence of stable MMs.

While, as mentioned above, SV plays the role of GS at $\gamma \leq 1$, the
MM represents GS at $\gamma \geq 1$. The \textit{switch} between SV and MM
at $\gamma =1$ is explained by comparison of values of the Hamiltonian for
SV and MM at equal values of $N$, which shows that SV and MM realize the
energy minimum at $\gamma \leq 1$ and $\gamma \geq 1$, respectively.
Accordingly, SV and MM are unstable, severally, at $\gamma >1$ and $\gamma <1
$: in these cases, SV starts spontaneous motion, while MM spontaneously
breaks the symmetry between its two components, tending to rearrange itself
into SV \cite{Sakaguchi14}. Furthermore, the analysis can be extended to $%
\gamma <0$ in Eqs. (\ref{GPERD1}) and (\ref{GPERD2}), i.e., to the system
combining self-attraction and cross-repulsion of the two components, in
which stable SVs exist as well \cite{Raymond}. In the same vein, one can
consider the system with competing self-repulsion and cross-attraction,
which maintains stable MMs \cite{Raymond}.

In addition to the two fundamental species of the 2D solitons, SV and MM,
Eqs. (\ref{GPERD1}) and (\ref{GPERD2}) give rise to excited states, obtained
by adding vorticity $M\geq 1$ to both components. In particular, the
excited-state variety of SV is supplied by an exact ansatz which is a
straightforward extension of (\ref{frf}), with $\phi _{+}\left( x,y,t\right)
=e^{-i\mu t+iM\theta }r^{m}f_{1}(r^{2}),~\phi _{-}\left( x,y,t\right)
=e^{-i\mu t+i\left( M+1\right) \theta }r^{m+1}f_{2}(r^{2})$, inputs
generating excited states of MM being more complex. However, all the excited
states, on the contrary to their fundamental counterparts, are completely
unstable \cite{Sakaguchi14}.

\textit{Motion and collisions of SVs and MMs.} Generation of moving 2D
solitons from the quiescent ones considered above is a nontrivial issue,
because SOC terms break the Galilean invariance of Eqs. (\ref{GPERD1}) and (%
\ref{GPERD2}). In particular, the application of the formal Galilean
transform for motion along the $y$ axis with velocity $v_{y}$, $\phi _{\pm
}\left( x,y,t\right) \equiv \tilde{\phi}_{\pm }(x,\tilde{y}\equiv
y-v_{y}t,t)\exp \left( iv_{y}y-iv_{y}^{2}t/2\right) $, casts Eqs. (\ref%
{GPERD1}) and (\ref{GPERD2}) with $\lambda _{D}=\Omega =0$ into the form
differing from the original one by the presence of linear mixing terms with
coefficient $\lambda v_{y}$:
\begin{eqnarray}
i\frac{\partial \phi _{+}}{\partial t} &=&-\frac{1}{2}\tilde{\nabla}^{2}\phi
_{+}-(|\phi _{+}|^{2}+\gamma |\phi _{-}|^{2})\phi _{+}+\lambda \tilde{D}%
^{[-]}\phi _{-}+\lambda v_{y}\phi _{-},  \notag \\
i\frac{\partial \phi _{-}}{\partial t} &=&-\frac{1}{2}\tilde{\nabla}^{2}\phi
_{-}-(|\phi _{-}|^{2}+\gamma |\phi _{+}|^{2})\phi _{-}-\lambda \tilde{D}%
^{[+]}\phi _{+}+\lambda v_{y}\phi _{+}  \label{mix}
\end{eqnarray}%
(the tilde in Eq. (\ref{mix}) implies the replacement of $\partial /\partial
y$ by $\partial /\partial \tilde{y}$). Stationary solutions to Eq. (\ref{mix}%
) are tantamount to steadily moving modes in the laboratory reference frame.
It was thus found that Eqs. (\ref{GPERD1}) and (\ref{GPERD2}) with $\lambda
_{D}=\Omega =0$ cannot generate moving solitons with $v_{x}\neq 0$, while MM
can be set in the steady motion in the direction of $y$, with the velocity
up to a certain maximum value, $|v_{y}|<v_{\max }$. Fixing the norm of the
moving MM, one observes that its amplitude decays with the increase of $v_{y}
$, vanishing at $v_{y}=v_{\max }$, i.e., the MM suffers delocalization at
this point \cite{Sakaguchi14}. For instance, $v_{\max }\approx 1.8$ for $%
\lambda =1$, $\gamma =2$ and $N=3.1$. SV may also be set in motion along $y$%
, but its limit velocity is very small (roughly, smaller by a factor $\sim 50
$ in comparison with MMs). Finally, the availability of MMs moving with
velocities $\pm v_{y}$ makes it possible to simulate collisions between
them, resulting in their fusion into a single mode, also of the MM type \cite%
{Sakaguchi14}.

\textit{Effect of combined Rashba and Dresselhaus SOC}. When the Dresselhaus
terms, with $\lambda _{D}\neq 0$, are present in Eqs. (\ref{GPERD1}) and (%
\ref{GPERD2}), an exact ansatz similar to one (\ref{frf}) is not available,
but a numerical SV solution can be constructed starting from initial guess $%
\phi _{+}^{(0)}\left( x,y\right) =A_{1}\exp \left( -\alpha _{1}r^{2}\right) $%
,$~\phi _{-}^{(0)}\left( x,y\right) =A_{2}e^{i\theta }r\exp \left( -\alpha
_{2}r^{2}\right) $, with constants $A_{1,2}$ and $\alpha _{1,2}>0$, while a
solution for MM is generated by the same ansatz (\ref{MM}) as above. As a
result, it is found that, again, SV and MM realize GS at $\gamma <1$ and $%
\gamma >1$, respectively \cite{we2}.

\begin{figure}[tbh]
\begin{center}
\includegraphics[width=0.45\textwidth]{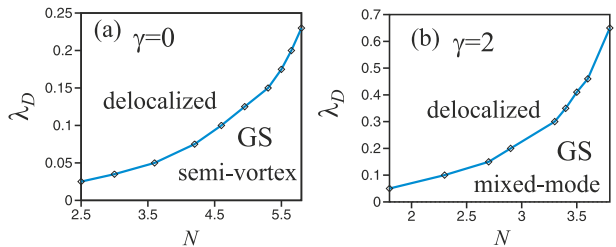}
\end{center}
\caption{(Color online) (a) and (b): Existence domains for the ground state
in the form of SV (with $\protect\gamma =0$) and MM ($\protect\gamma =2$),
respectively, in the plane of norm $N$ and Dresselhaus coupling constant $%
\protect\lambda _{D}$. Other parameters in Eqs. (\protect\ref{GPERD1}) and (%
\protect\ref{GPERD2}) are $\protect\lambda =1$ and $\Omega =0$. The plots
are borrowed from \protect\cite{we2} and \protect\cite{Romanian}.}
\label{fig3}
\end{figure}

The most essential effect caused by the inclusion of the Dresselhaus terms
is destruction (delocalization) of SV and MM when $\lambda _{D}\ $exceeds
certain critical values, which are growing functions of $N$ and depend on $%
\gamma $. In other words, SV and MM exist in intervals, respectively, $%
N_{\min }^{(\mathrm{SV})}(\lambda _{D})<N<N_{\mathrm{T}}$ and $N_{\min }^{(%
\mathrm{MM})}(\lambda _{D})<N<2\left( 1+\gamma \right) ^{-1}N_{\mathrm{T}}$,
where $N_{\mathrm{T}}$ is, as above, the TS norm, cf. Eq. (\ref{N<N}) \cite%
{we2}. These results are summarized in Fig. \ref{fig3}, which shows
stability regions for SVs and MMs in the $\left( N,\lambda _{D}\right) $
plane at $\gamma =0$ and $\gamma =2$, respectively. The presence of the
threshold value $N_{\min }$ necessary for the existence of the solitons is a
drastic difference from the system with $\lambda _{D}=0$, cf. Figs. \ref%
{fig1}(b) and \ref{fig2}(b). On the other hand, it is easy to see that the
upper limits, $N_{\mathrm{T}}$ and $2\left( 1+\gamma \right) ^{-1}N_{\mathrm{%
T}}$, do not depend on $\lambda _{D}$. Note that the critical values of the
Dresselhaus coupling constant, up to which the solitons persist, are
essentially larger for MM than for SV.

\textit{Effects of the Zeeman splitting (ZS)}. Through\ the trend to
populate one pseudo-spin component and depopulate the other one, the ZS
essentially affects the SOC soliton phenomenology \cite{we2}. In particular,
the stationary version of Eqs. (\ref{GPERD1}) and (\ref{GPERD2}), with $\phi
_{\pm }=e^{-i\mu t}u_{\pm }\left( x,y\right) $, admits the application of a
simple analytical approximation in the limit of large $\Omega $, setting $%
\mu =-\Omega +\delta \mu $, with$~\left\vert \delta \mu \right\vert \ll
\Omega $. In this limit, fixing $\lambda =1$ and $\lambda _{D}=0$, one can
eliminate the depopulated component in favor for the other one: $%
u_{-}\approx \left( 2\Omega \right) ^{-1}D^{[+]}u_{+}$, which obeys the
stationary NLSE, $\left( \delta \mu \right) u_{+}=-(1/2)\left( 1-\Omega
^{-1}\right) \nabla ^{2}u_{+}-|u_{+}|^{2}u_{+}$. Up to rescaling, this NLSE
gives rise to the zero-vorticity TS, while $u_{-}$ is a small vortex
component of the SV complex. With regard to the smallness of $1/\Omega $,
the SV's norm is found, by means of the scaling argument, as%
\begin{equation}
N=\left( 1-\Omega ^{-1}\right) N_{\mathrm{T}}+\mathcal{O}\left( \Omega
^{-2}\right) .  \label{NT}
\end{equation}%
Thus, the norm is slightly smaller than the collapse threshold, keeping the
SV protected against the collapse. The conclusion is that, at large $\Omega $%
, GS is of the SV type, irrespective of the value of the cross-attraction
coefficient, $\gamma $, which does not appear in this approximation.

\begin{figure}[tbh]
\par
\begin{center}
\subfigure[]{\includegraphics[width=0.25\textwidth]{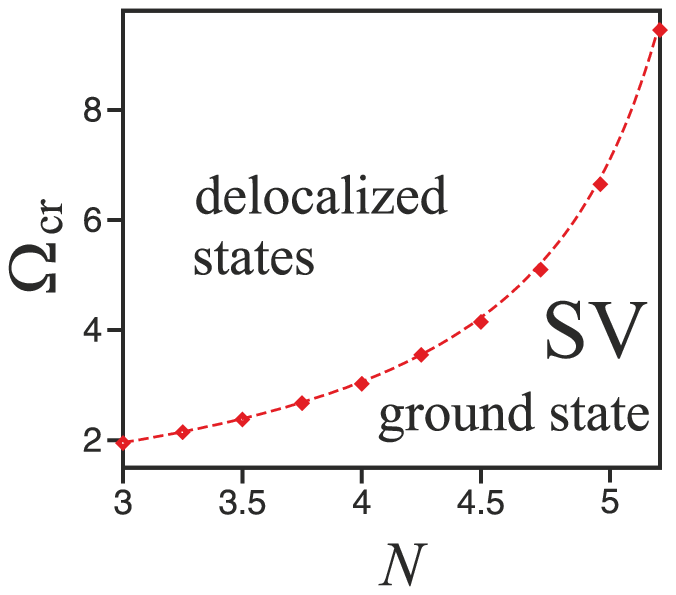}}%
\subfigure[]{\includegraphics[width=0.25\textwidth]{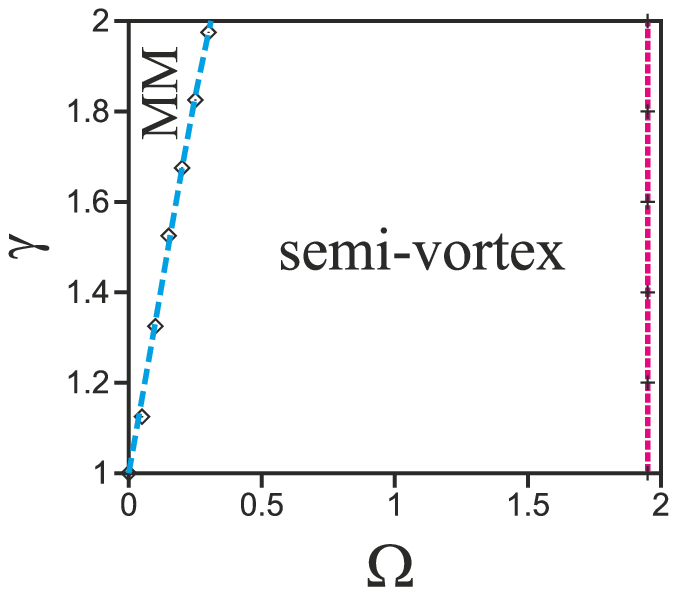}}
\end{center}
\caption{(Color online) (a) The critical ZS strength, $\Omega _{\mathrm{cr}}$%
, at which GS in the form of SV suffers delocalization, vs. $N$, for $%
\protect\lambda =1$ and $\protect\gamma =\protect\lambda _{D}=0$ in Eqs. (%
\protect\ref{GPERD1}) and (\protect\ref{GPERD2}). (b) The numerically found
boundary, in the $(\Omega ,\protect\gamma )$ plane, for fixed $N=3$ and $%
\protect\lambda =1$, $\protect\lambda _{D}=0$, between GSs of the MM and SV
types.The vertical line corresponds to $\Omega \approx 1.95$ at which the SV
suffers the delocalization. The plots are borrowed from \protect\cite{we2}
and \protect\cite{Romanian}.}
\label{fig4}
\end{figure}

Because Eq. (\ref{NT}) produces only values of $N$ close to $N_{\mathrm{T}}$
at large $\Omega $, one may expect that SVs with smaller norms suffer
delocalization with the increase of $\Omega $ at some critical value $\Omega
_{\mathrm{cr}}$. This is indeed demonstrated by both the variational
approximation and numerical results \cite{we2}, see Fig. \ref{fig4}(a) (for
instance, $\Omega _{\mathrm{cr}}(\gamma =0,N=3)\approx 1.95$). In agreement
with the above approximation, at $\Omega <\Omega _{\mathrm{cr}}$ stable SV
keeps vorticity $m=1$ in the component with a smaller norm, and zero
vorticity in the heavier component. As concerns MM, it is ousted by SV with
the increase of $\Omega $. This is seen in the increase of the value of $%
\gamma _{\mathrm{cr}}$, above which MM plays the role of GS: while, as
mentioned above, $\gamma _{\mathrm{cr}}=1$ is the universal SV-MM boundary
at $\Omega =0$, Fig. \ref{fig4}(b) demonstrates that $\gamma _{\mathrm{cr}}$
grows nearly linearly with the increase of $\Omega $. At $\gamma >$ $\gamma
_{\mathrm{cr}}$, the symmetry between two MM's components, which holds at $%
\Omega =0$, is broken, the norm of $\phi _{+}$ being larger, with a
progressively smaller vortex part in it.

\begin{figure}[tbh]
\centering\includegraphics[width=0.4\columnwidth]{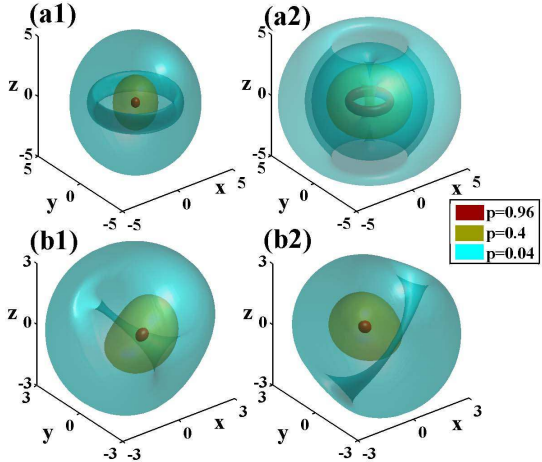}
\caption{(Color online) Density profiles of metastable 3D solitons with $N_{%
\mathrm{3D}}=8$, of the SV (a) and MM (b) types, generated by Eq. (\protect
\ref{GPE}) with $\protect\lambda =1$ and $\protect\gamma =0.3$ in (a), or $%
\protect\gamma =1.5$ in (b). The SV's zero-vorticity and vortex components, $%
|\protect\phi _{+}|$ and $|\protect\phi _{-}|$, are plotted in (a1) and
(a2), respectively, while the MM's components are shown in (b1) and (b2). In
each panel, colors represent surfaces of fixed absolute values, $|\protect%
\phi _{\pm }|=p|\protect\phi _{\pm }|_{\mathrm{max}}$, with $p=0.96$, $0.40$%
, and $0.04$. The plots are borrowed from \protect\cite{HP}.}
\label{fig5}
\end{figure}

\textit{Metastable 3D solitons}. As said above, in 2D the SOC\ system
creates the otherwise missing GS at $N<N_{\max }$, while inputs with $%
N>N_{\max }$ undergo the collapse. As shown in \cite{HP}, SOC added to the 3D
system (\ref{GPE}) can never suppress the supercritical collapse, therefore
the 3D system has no GS, which would realize an absolute minimum of the
energy. Nevertheless, a variational analysis predicts metastable 3D solitons
which represent a local energy minimum, i.e., such solitons are stable
against small perturbations. Numerically, they can be produced by inputs
similar to those adopted in the 2D system (see, e.g., Eq. (\ref{MM})), with
an extra factor $\exp \left( -\gamma z^{2}\right) $ providing the
localization in $z$ for $\gamma >0$. As a result, both the variational
approximation and numerical analysis generate families of 3D metastable
solitons, which are SV and MM at $\gamma <1$ and $\gamma >1$, respectively,
see examples in Fig. \ref{fig5}. Fixing $\lambda =1$ in Eq. (\ref{GPE}), the
conclusion is that, similar to the 2D setting (cf. Eq. (\ref{N<N})), the
metastable 3D solitons exist at values of the 3D norm $N_{\mathrm{3D}%
}<\left( N_{\mathrm{3D}}\right) _{\max }(\gamma )$, the largest value being $%
\left( N_{\mathrm{3D}}\right) _{\max }(\gamma =0)\approx 11.5$ \cite{HP}.

\begin{figure}[t]
{\includegraphics[width=0.7\columnwidth]{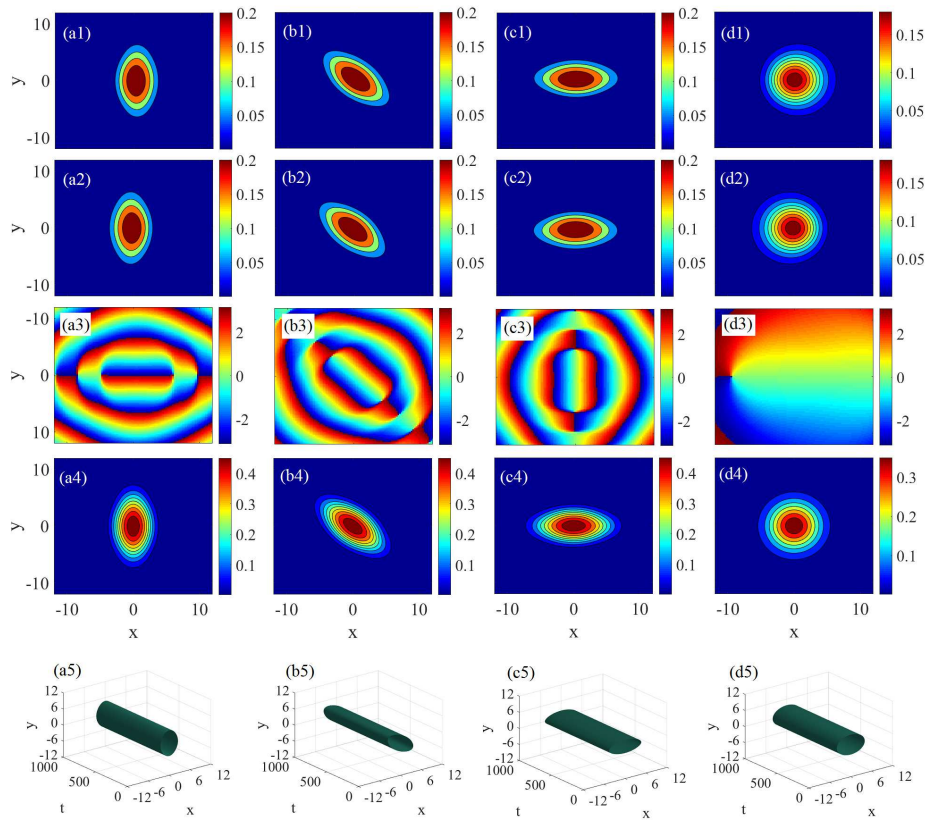}}
\caption{(Color online) Typical examples of elongated QDs (\textquotedblleft
quantum droplets"), generated by Eq. (\protect\ref{GPE2}) with $(g,\protect%
\lambda ,N)=(2,1,20)$, are displayed in columns (a-c), with the vertical,
diagonal, and horizontal orientations, respectively. The first and second
rows display density patterns $|\protect\phi _{+}(\mathbf{r})|^{2}$ and $|%
\protect\phi _{-}(\mathbf{r})|^{2}$, the phase pattern of $\protect\phi _{+}$
is presented in the third row, the fourth row shows the total density
profile, $|\protect\phi _{+}(\mathbf{r})|^{2}+|\protect\phi _{-}(\mathbf{r}%
)|^{2}$, and the fifth row illustrates the stability of the QDs in direct
simulations of Eq. (\protect\ref{GPE2}) (shown by the evolution of the total
density). Column (d) is an example of QD with $(g,\protect\lambda %
,N)=(2,0.2,20)$, which is nearly isotropic, as $\protect\lambda $ is small.
The plots are borrowed from \protect\cite{Raymond}.}
\label{fig6}
\end{figure}

\textit{2D\textquotedblleft quantum droplets" (QDs) of the MM type
stabilized by the LHY corrections. }A general property of absolutely stable
2D solitons and metastable 3D ones is that they exist below a critical value
of the norm, see, e.g., Eq. (\ref{N<N}). On the other hand, as LHY terms may
stabilize 3D and 2D solitons against the collapse, in the form of QDs \cite%
{Petrov,Grisha}, the SOC system (\ref{GPE2}), including the LHY terms in
their 2D form,\ gives rise to a family of stable QDs of the MM type, without
any upper boundary $N_{\max }$. As shown in Fig. \ref{fig6}, a
characteristic feature of the SOC-affected QDs is their elongated shape,
which may have arbitrary orientation in the $\left( x,y\right) $ plane, due
to the azimuthal invariance of Eqs. (\ref{GPE}): if there is a stationary
solution $\left\{ \phi _{+}\left( r,\theta \right) ,\phi _{-}\left( r,\theta
\right) \right\} $, its rotated version, $\left\{ \tilde{\phi}_{+},\tilde{%
\phi}_{-}\right\} =\left\{ \phi _{+}\left( r,\theta +\theta _{0}\right)
,~e^{-i\theta _{0}}\phi _{-}\left( r,\theta +\theta _{0}\right) \right\} $,
with arbitrary angle $\theta _{0}$, is a solution too \cite{Raymond}.
Remarkably, all QDs are stable, as illustrated by plots in the bottom row of
Fig. \ref{fig6}. Lastly, if SOC terms of both the Rashba and Dresselhaus
types are included, Eq. (\ref{GPE2}) generates QDs for any ratio $\lambda
_{D}/\lambda $ of their strengths, including $\lambda _{D}/\lambda =1$,
while, in the absence of the LHY effect, MM exists only at essentially
smaller values of $\lambda _{D}/\lambda $, see Fig. \ref{fig3}(b).

\section{Conclusion}

This mini-review focuses on recently produced theoretical results which
demonstrate the possibility of the creation of absolutely stable 2D \cite%
{Sakaguchi14,we2} and metastable 3D \cite{HP} solitons supported by SOC in
binary BEC with attractive nonlinearity. The most essential prediction is
that, on the contrary to the commonly known instability of 2D solitons and
solitary vortices created by cubic self-attraction, two \emph{stable}
soliton species, SV (semi-vortex) and MM (mixed mode), are supported by SOC.
In 2D, SV and MM represent the GS (ground state) when self-attraction in
each component is, respectively, stronger or weaker than its
cross-interaction counterpart. SV is made more favorable if ZS (Zeeman
splitting) is applied. A broad class of stable MMs was recently predicted
\cite{Raymond} in the 2D system which includes beyond-mean-field (LHY)
corrections. These results suggest novel possibilities for the creation of
stable vorticity-bearing solitons in BEC. An analog of SOC can also be
implemented in optics, which was used to predict stable spatiotemporal
solitons in nonlinear dual-core planar waveguides \cite{optics,NJP}.

The work can be extended in other directions. In particular, if SOC acts in
a confined 2D or 3D spatial domain (similar to the 1D setting introduced in
\cite{Konotop}), it is relevant to identify a minimum size of the domain
which is sufficient for the creation of stable solitons. Further, it is
interesting to construct 3D solitons in the SOC system which includes LHY
terms, and to consider effects of ZS on 3D solitons. It may also be relevant
to apply the concept of the \textquotedblleft nonlinearity management" \cite%
{Springer} to the SOC system, periodically switching it, by means of
time-dependent Feshbach resonance, between settings in which SV and MM
represent GS.

\textit{Acknowledgments}. I appreciate valuable collaborations with H.
Sakaguchi, B. Li, E. Ya. Sherman, and Y. Li, which have produced results
surveyed in this mini-review. I thanks G. Muga for the invitation to write
this article. My work on the topic is partly supported by Israel Science
Foundation, project No. 1287/17.


\begin{thebibliography}{99}
\bibitem{BEC-book1} L. P. Pitaevskii and S. Stringari, \textit{Bose-Einstein
Condensation}, Oxford University Press, Oxford, 2003.

\bibitem{Bagnato} V. S. Bagnato, D. J. Frantzeskakis, P. G. Kevrekidis, B.
A. Malomed, and D. Mihalache, Rom. Rep. Phys. \textbf{67}, 5 (2015).

\bibitem{emulator} P. Hauke, F. M. Cucchietti, L. Tagliacozzo, I. Deutsch,
and M. Lewenstein, Rep. Prog. Phys. \textbf{75}, 082401 (2012).

\bibitem{Malomed} B. Malomed, L. Torner, F. Wise, and D. Mihalache, J. Phys.
B: At. Mol. Opt. Phys. \textbf{49}, 170502 (2016).

\bibitem{KA} Y. S. Kivshar and G. P. Agrawal, \textit{Optical Solitons: From
Fibers to Photonic Crystals} (Academic Press, San Diego, 2003).

\bibitem{Nature} Y. J. Lin, K. Jimenez-Garcia, and I. B. Spielman, Nature
\textbf{471}, 83 (2011).

\bibitem{Anderson} B. M. Anderson, G. Juzeli\={u}nas, V. M. Galitski, and I.
B. Spielman, Phys. Rev. Lett. \textbf{108}, 235301 (2012).

\bibitem{Spielman} I. B. Spielman, Ann. Rev. Cold At. Mol. \textbf{1}, 145
(2012).

\bibitem{NatureRev} V. Galitski and I. B. Spielman, Nature \textbf{494}, 49
(2013).

\bibitem{Gedeminas} N. Goldman, G. Juzeliunas, P. \"{O}hberg, and I. B.
Spielman, Rep. Progr. Phys. \textbf{77}, 126401 (2014).

\bibitem{Zhai} H. Zhai, Rep. Prog. Phys. \textbf{78}, 026001 (2015).

\bibitem{Campbell} D. L. Campbell, G. Juzeli\={u}nas, and I. B. Spielman,
Phys. Rev. A \textbf{84}, 025602 (2011).

\bibitem{Dresselhaus} G. Dresselhaus, Phys. Rev. \textbf{100}, 580 (1955).

\bibitem{Rashba} Y. A. Bychkov and E. I. Rashba, J. Phys. C \textbf{17},
6039 (1984).

\bibitem{2D-experiment} Z. Wu, L. Zhang, W. Sun, X.-T. Xu, B.-Z. Wang, S.-C.
Ji, Y. Deng, S. Chen, X.-J. Liu, and J.-W. Pan, Science \textbf{354}, 83
(2016).



\bibitem{Sakaguchi14} H. Sakaguchi, B. Li, and B. A. Malomed, Phys. Rev. E
\textbf{89}, 032920 (2014).

\bibitem{we2} H. Sakaguchi, E. Ya. Sherman, and B. A. Malomed, {Phys. Rev. E}
\textbf{94}, 032202 (2016).

\bibitem{Fukuoka2} H. Sakaguchi and B. A. Malomed, Phys. Rev. E \textbf{90},
062922 (2014).

\bibitem{Cardoso} L. Salasnich, W. B. Cardoso, and B. A. Malomed, Phys. Rev.
A \textbf{90}, 033629 (2014).

\bibitem{gap-sol} V. E. Lobanov, Y. V. Kartashov, and V. V. Konotop, Phys.
Rev. Lett. \textbf{112}, 180403 (2014).

\bibitem{NJP} H. Sakaguchi and B.A. Malomed, New J. Phys. \textbf{18},
105005 (2016).

\bibitem{HP} Y.-C. Zhang, Z.-W. Zhou, B. A. Malomed, and H. Pu, Phys. Rev.
Lett. \textbf{115}, 253902 (2015).

\bibitem{vortex1} T. Kawakami, T. Mizushima, and K. Machida, Phys. Rev. A
\textbf{84}, 011607 (2011).

\bibitem{vortex2} H. Sakaguchi and B. Li, Phys. Rev. A \textbf{87}, 015602
(2013).


\bibitem{rf:2} B. Ramachandhran, B. Opanchuk, X.-J. Liu, H. Pu, P. D.
Drummond, and H. Hu, Phys. Rev. A \textbf{85}, 023606 (2012).

\bibitem{Fetter} A. L. Fetter, Rev. Mod. Phys. \textbf{81}, 647 (2009).

\bibitem{skyrmions2} T. Kawakami, T. Mizushima, M. Nitta, and K. Machida,
Phys. Rev. Lett. \textbf{109}, 015301 (2012).

\bibitem{Cornish} S. L. Cornish, N. R. Claussen, J. L. Roberts, E. A.
Cornell, and C. E. Wieman, Phys. Rev. Lett. 85, 1795 (2000).

\bibitem{Feshbach} C. Chin, R. Grimm, P. Julienne, and E. Tiesinga, Rev.
Mod. Phys. \textbf{82}, 1225 (2010).

\bibitem{Randy} K. E. Strecker, G. B. Partridge, A. G. Truscott, and R. G.
Hulet, Nature \textbf{417}, 150 (2002).

\bibitem{Salomon} L. Khaykovich, F. Schreck, G. Ferrari, T. Bourdel, J.
Cubizolles, L. D. Carr, Y. Castin, and C. Salomon, Science \textbf{296},
1290 (2002).

\bibitem{Weiman} S. L. Cornish, S. T. Thompson, and C. E. Wieman, Phys. Rev.
Lett. \textbf{96}, 170401 (2006).

\bibitem{Berge} L. Berg\'{e}, Phys. Rep. \textbf{303}, 259 (1998).

\bibitem{Fibich} G. Fibich, \textit{The Nonlinear Schr\"{o}dinger Equation:
Singular Solutions and Optical Collapse} (Springer: Heidelberg, 2015).

\bibitem{Mardonov15} Sh. Mardonov, E. Ya. Sherman, J. G. Muga, H.-W. Wang,
Y. Ban, and X. Chen, Phys. Rev. A \textbf{91}, 043604 (2015).

\bibitem{review} B. A. Malomed, D. Mihalache, F. Wise, and L. Torner, J.
Optics B: Quant. Semicl. Opt. \textbf{7}, R53 (2005).

\bibitem{me} B. A. Malomed, Eur. Phys. J. Special Topics \textbf{225}, 2507
(2016).

\bibitem{Mihalache} D. Mihalache, Rom. Rep. Phys. \textbf{69}, 403 (2017).

\bibitem{Raymond} Y. Li, Z. Luo, Y. Liu, Z. Chen, C. Huang, S. Fu, H. Tan,
and B. A. Malomed, New. J. Phys. \textbf{19}, 113043 (2017).

\bibitem{LHY} T. D. Lee, K. Huang, and C. N. Yang, Phys. Rev. \textbf{106},
1135 (1957).

\bibitem{Petrov} D. S. Petrov, Phys. Rev. Lett. \textbf{115}, 155302 (2015).

\bibitem{Grisha} D. S. Petrov and G. E. Astrakharchik, Phys. Rev. Lett.
\textbf{117}, 100401 (2016).

\bibitem{Let1} C. R. Cabrera, L. Tanzi, J. Sanz, B. Naylor, P. Thomas, P.
Cheiney, and L. Tarruell, Science \textbf{359}, 301 (2018)

\bibitem{Let2} P. Cheiney, C. R. Cabrera, J. Sanz, B. Naylor, L. Tanzi, and
L. Tarruell, Phys. Rev. Lett. \textbf{120}, 135301 (2018).

\bibitem{Ing} G. Semeghini, G. Ferioli, L. Masi, C. Mazzinghi, L. Wolswijk,
F. Minardi, M. Modugno, G. Modugno, M. Inguscio, and M. Fattori,
arXiv:1710.10890.

\bibitem{Romanian} H. Sakaguchi, B. Li, E. Ya. Sherman, and B. A. Malomed,
Romanian Rep. Phys. \textbf{70}, 502 (2018).

\bibitem{VaKo} M. Vakhitov and A. Kolokolov, Radiophys. Quant. Electron.
\textbf{16}, 783 (1973).

\bibitem{Townes} R. Y. Chiao, E. Garmire, and C. H. Townes, Phys. Rev. Lett.
\textbf{13}, 479 (1964).

\bibitem{optics} Y. V. Kartashov, B. A. Malomed, V. V. Konotop, V. E.
Lobanov, and L. Torner, Opt. Lett. \textbf{40}, 1045 (2015).

\bibitem{Konotop} Y. V. Kartashov, V. V. Konotop, and D. A. Zezyulin, Phys.
Rev. A \textbf{90}, 063621 (2014).

\bibitem{Springer} B. A. Malomed, \textit{Soliton Management in Periodic
Systems} (Springer: New York, 2006).
\end{thebibliography}
\end{document}